\documentclass[aps,pre,twocolumn,showpacs,floatfix]{revtex4}

\usepackage{graphicx}
\usepackage{natbib}

\usepackage{amsmath,bm,epsfig}
\usepackage{epsfig,color}
\usepackage{bm}

\renewcommand{\it}[1]{\textit{#1}}

\begin{document}
\setcounter{page}{1}

\title{Heat transport by laminar
boundary layer flow with polymers}
\author{Roberto Benzi$^1$, Emily S.C. Ching$^{2,3}$ and Vivien W.S. Chu$^2$}
\affiliation{$^1$ Dip. di Fisica and INFN, Universit\`a ``Tor
Vergata", Via della Ricerca Scientifica 1, I-00133 Roma, Italy \\
$^2$ Department of Physics,
The Chinese University of Hong Kong, Shatin, Hong Kong \\
$^3$ Institute of Theoretical Physics, The Chinese University of
Hong Kong, Shatin, Hong Kong}

\date{\today}

\begin{abstract}
{Motivated by recent experimental observations,
we consider a steady-state Prandtl-Blasius boundary layer flow
with polymers above a slightly heated horizontal plate and study how the
heat transport might be affected by the polymers. We discuss how a
set of equations can be derived for the problem and how these equations
can be solved numerically by an iterative scheme.
By carrying out such a scheme, we find that the effect of the polymers
is equivalent to producing a space-dependent effective viscosity that
first increases from the zero-shear value at the plate
then decreases rapidly back to the zero-shear value far from the plate. We
further show that such an effective viscosity leads to an enhancement in
the drag, which in turn leads to a reduction in heat transport.}

\end{abstract}
\maketitle

\section{Introduction}
\label{introduction}

It has been known for more than 60 years that adding polymers into
turbulent wall-bounded flows can reduce the friction drag
significantly (see, for example, \cite{Sreeni,RMP} and references
therein). This effect of polymer additives on mass transport has
been studied extensively during the past 60 or so years. On the
other hand, the effect of polymers additives on heat transport is
much less studied. Recently, an experimental study
reported~\cite{AhlersPRL} that for turbulent Rayleigh-B{\'e}nard
(RB) convection of water, confined within a cylindrical cell
heated below and cooled on top, adding polymers to the flow
reduces the heat transport. In turbulent RB convection within a
container of given shape, the flow state is characterized by two
dimensionless parameters: the Rayleigh number (Ra) and the Prandtl
number (Pr), where Ra measures the size of the thermal forcing and
Pr is the ratio of the kinematic viscosity to the thermal
diffusivity of the fluid. Moreover, there is an exact
balance~\cite{SiggiaReview} between the heat transport and the
energy and thermal dissipation rates. The energy and thermal
dissipation rates can be decomposed as sums of contributions from
the bulk of the flow and from the boundary layers near the top and
bottom plates~\cite{GLtheory}. The experimental study
reported~\cite{AhlersPRL} was conducted at a Ra of the order of
$10^{10}$. At such a moderate Ra, the energy and thermal
dissipation rates are mostly due to the contribution from the
boundary layers~\cite{GLtheory}. This suggests that the observed
reduction in heat transport is likely to be an effect of the
polymers interacting with the boundary layer flow in turbulent RB
convection. Experimental measurements~\cite{XiaJFM} indicated that
the average velocity and temperature boundary layer profiles in
turbulent RB convection at moderate Ra could be described by those
profiles in the steady-state Prandtl-Blasius boundary layer
flow~\cite{Landau,Schlichting} above a slightly heated flat plate.

Motivated by these experimental observations, we
study a steady-state Prandtl-Blasius boundary layer flow with polymers
near a slightly heated plate and focussed particularly on
the possible effect of the polymer additives on the heat transport.
Physically, we can think of this boundary layer flow as the flow near
the bottom plate of the convection cell in turbulent RB convection.

This paper is organized as follows. In Section~\ref{problem}, we formulate the
problem and set up the equations of motion for the system. We discuss how the
set of equations can be solved numerically by an iterative
scheme in Section~\ref{cal}. After solving the problem using such a scheme,
our results show that the effect of the polymers
is equivalent to producing a space-dependent effective viscosity that
first increases from the zero-shear value at the plate
then decreases rapidly back to the zero-shear value far from the plate.
We further show that such an effective viscosity would lead to an enhancement
in the friction drag and a reduction in heat transport.
We shall present and discuss our results in
Section~\ref{result}. Finally, we shall summarize and conclude in
Section~\ref{conclusion}.

\section{The Problem}
\label{problem}

For the Prandtl-Blasius boundary layer flow above a large flat
plate, the velocity equation is:
\begin{equation}
\label{PB}
v_x \partial_x v_x + v_y \partial_y v_x = \nu \partial^2_{yy} v_x
\end{equation}
Here $x$ denotes the direction along the plate,
$y$ denotes the direction away from
the plate, and $\nu$ is the kinematic viscosity of the fluid.
Both $v_x$ and $v_y$ vanish at the plate and $v_x$ merges with the
uniform mainstream velocity $U$ far away from the plate.
The crucial point about Eq.~(\ref{PB}) is that the viscous term
is balanced against the nonlinear advection term
and that the flow
changes more rapidly away from the plate than along the plate such that
$\partial_y \gg \partial_x$.
The latter condition is satisfied for flows with large
Reynolds number.

Introducing the variable
\begin{equation}
\xi \equiv \sqrt{\frac{U}{\nu x}} y
\label{xi}
\end{equation}
and the stream function
\begin{equation}
\label{stream}
\Psi(x,y) \equiv  \sqrt{\nu x U} \phi(\xi)
\end{equation}
such that
\begin{equation}
v_x = \partial_y \Psi \ ; \qquad
v_y = -\partial_x \Psi
\end{equation}
we obtain the famous Blasius equation~\cite{Landau}
\begin{equation}
2 \phi_{\xi \xi \xi } + \phi \phi_{\xi \xi} = 0
\label{BPphi}
\end{equation}
showing that the velocity profile is self-similar
in the dimensionless variable $\xi$
at different position $x$ along the plate.
Here $\phi_{\xi}$ denotes $\partial_{\xi} \phi$.
The boundary conditions are
\begin{equation}
\label{velBC}
\phi(0) = \phi_{\xi} (0) = 0  \ ; \qquad \phi_{\xi}(\infty)=1
\end{equation}
as $v_x \to U$ when it is far away from the plate.
The plate is slightly heated at a temperature $T_1$ above the
ambient temperature $T_0$ far away from the plate.
Writing the temperature field as
\begin{equation}
T(x,y) = T_0 + (T_1-T_0)\theta(\xi)
\end{equation}
[Relating to turbulent RB convection, $T_0$ can be taken as the
temperature at the center of the cell and thus the total temperature difference
across the cell is $2(T_1-T_0)$]. Then $\theta$ satisfies the equation
\begin{equation}
\label{temperature} 2 \theta_{\xi \xi} + \phi \theta_{\xi} {\rm Pr} = 0
\end{equation}
where ${\rm Pr} = \nu / \kappa $ and $\kappa$ is the thermal diffusivity
of the fluid, and the boundary conditions are:
\begin{equation}
\label{tempBC}
\theta(0)=1  \ ; \qquad \theta(\infty)=0
\end{equation}

We want to investigate the effect of polymers on heat transport in
this laminar Prandtl-Blasius flow. The polymers produces an
additional stress in the momentum equation of the fluid. This
polymer stress ${\cal T}_{ij}$ depends on the amount of stretching
of the polymers and is thus a function of the dimensionless
conformation tensor $R_{ij}$ of the polymers. Let the vector
$\vec{d}$ denote the polymer end-to-end distance and $\rho_0$ be
the polymer radius in the unstretched regime, then $R_{ij}$
represents the average over many ($N$) polymers in a small region
around the point $(x,y)$ of the product $d_i d_j/\rho_0^2$, i.e.
$R_{ij} = N^{-1} \Sigma d_i d_j/\rho_0^2$. In the simplest
Oldroyd-B model of polymers~\cite{Bird},
\begin{equation}
{\cal T}_{ij} = \frac{\nu_p}{\tau} \left( R_{ij} - \delta_{ij} \right)
\label{OldroydB}
\end{equation}
where $\nu_p$ is the polymer contribution
to the viscosity of the solution at zero shear and $\tau$ is the
relaxation time of the polymers.
Thus in the presence of polymers, the equation of motion for the
velocity field is
modified by an additional stress that
depends on $R_{ij}$. By employing the same ideas
leading to the Blasius equation [Eq.~(\ref{PB})] we have
\begin{equation}
\label{PBR} v_x \partial_x v_x + v_y \partial_y v_x = \nu
\partial^2_{yy} v_x + \partial_y \left[\frac{\nu_p}{\tau} R_{xy} \right]
\end{equation}
It has been shown that the relaxation time of the polymers can be
significantly increased by the stretching of the
polymers~\cite{VincenziPRL}. To model this effect, we let
\begin{equation}
\tau = \tau_0 \left( \frac{1+a R}{1+a} \right)
\label{Dynamic_tau}
\end{equation}
where $\tau_0$ is the bare Zimm relaxation time,
$R \equiv (R_{xx}+R_{yy})^{1/2}$, and $a > 0$ is a parameter.
When there is no stretching, $R=1$ and $\tau$ reduces back to $\tau_0$.
When there is stretching, $R > 1$ and $\tau > \tau_0$.

In the presence of polymers, the transformation using $\xi$ does
not lead to a similarity solution in general in that explicit
appearance of $x$ remains in the equation for $\phi$. This is
known in the literature. Similarity solution has been obtained in
some special cases with certain velocity or temperature boundary
conditions~\cite{Olagunju2006,Bataller2008} that might not have
direct physical relevance. Here to circumvent this difficulty, we
recall that the Prandtl-Blasius flow is meant to be applicable
when $x$ is large (such that $\partial_y \gg \partial_x$). Thus we
make the following approximations:
\begin{eqnarray}
v_y &\approx& \frac{1}{2} \sqrt{\frac{\nu_0 U}{L}} (\xi \phi_\xi - \phi)
\label{approx1}\\
\partial_x &\approx& -\frac{\xi}{2L} \frac{d}{d\xi} \ ;
\qquad
\partial_y \approx \sqrt{\frac{U}{\nu_0 L}} \frac{d}{d\xi}
\label{approx2}
\end{eqnarray}
by putting $x=L$, the length of the (long) plate, and
replacing $\nu$ in $\xi$ [Eq.~(\ref{xi})] by
$\nu_0=\nu+\nu_p$, where $\nu_0$ is the total viscosity of the
polymer solution at zero shear. That is, in the presence of polymers,
we have
\begin{equation}
\xi = \sqrt{\frac{U}{\nu_0 x}} y \qquad {\rm with \ polymers}
\label{xinew}
\end{equation}
With these approximations,
the scaling transformation of $\xi$ leads to a
similarity solution. The resulting
modified Blasius equation is:
\begin{equation}
\label{basic} -\frac{1}{2} \phi \phi_{\xi\xi} = (1-\gamma)
\phi_{\xi \xi \xi } + \frac{\gamma}{{\rm Wi} \sqrt{\rm Re}}
\frac{d}{d \xi} \left[ \frac{(1+a)R_{xy}}{1+aR} \right]
\end{equation}
where the Weissenberg number (Wi) and the Reynolds number (Re) are
defined as
\begin{equation}
{\rm Wi}\equiv \frac{\tau_0 U}{L} \ , \qquad
{\rm Re} \equiv \frac{UL}{\nu_0}
\end{equation}
and $\gamma \equiv \nu_p /\nu_0$
is a function of the polymer concentration.
As usual in the Prandtl-Blasius approximation, all terms of the order
of 1/Re are nelgected in Eq.~(\ref{basic}).

We would like to study how the heat transport is affected by the polymers.
In turbulent RB convection, it is common to measure the heat flux $Q$
in terms of the
dimensionless Nusselt number (Nu), which is the ratio of $Q$ to that
when there is only conduction, defined by
\begin{equation}
{\rm Nu}=  \displaystyle \frac{Q}{2k(T_1-T_0)/H} = \displaystyle
\frac{ \langle \displaystyle -\frac{\partial T}{\partial y}
\bigg |_{y=0} \rangle_{A} }{2(T_1-T_0)/H}
\end{equation}
where $k$ is the thermal conductivity of the fluid,
$H$ is the height of the convection cell, and $\langle \ldots \rangle_A$
is the average over the cross section of the cell. For
the Prandtl-Blasius flow, taking $H=L$ and dropping the numerical factor,
Nu can be estimated as
\begin{equation}
{\rm Nu}= \sqrt{\frac{UL}{\nu_0}} \ [-\theta_{\xi}(0)]
\label{Nutheta}
\end{equation}

To proceed, we must supplement
Eq. (\ref{basic}) with a specific information on $R_{xy}$. In a fluid flow of
velocity $\vec{v}$, the components of the dimensionless
polymer end-to-end distance ${l_i}={d_i}/\rho_0$, $i=x,y$ obey the
differential equations:
\begin{equation}
\label{polymer} \frac{d l_i}{dt } = - \frac{1}{2 \tau} (l_i -
l_{0i}) + l_j \partial_j v_i + {\rm thermal \ noise}
\end{equation}
where $l_{0x}= \cos \alpha$ and $l_{0y}=\sin \alpha$ and $\alpha$ is a random
angle uniformly distributed in $[0,2\pi]$. 
Neglecting the thermal noise, we can rewrite
Eq.~(\ref{polymer})
for the two-dimensional Prandtl-Blasius flow as
\begin{eqnarray}
\label{polx} -\frac{L}{U}\frac{d l_x}{d t} &=&
\frac{(1+a)(l_x-l_{0x}) }{2 {\rm Wi}(1+aR)}  + \frac{\xi \phi_{\xi
\xi}}{2} l_x -
\sqrt{\rm Re} \phi_{\xi \xi} l_y \\
 - \frac{L}{U}\frac{d l_y}{d t} &=&
\frac{(1+a)(l_y-l_{0y})}{2 {\rm Wi}(1+aR)}  + \frac{\xi^2
\phi_{\xi\xi}}{4 \sqrt{\rm Re}} l_x  - \frac{\xi \phi_{\xi
\xi}}{2}  l_y \label{poly}
\end{eqnarray}
In order to obtain $l_i$ as a function of $\xi$, we assume that each
polymer follows the streamline of the flow such that
$d\xi/d t = v_x \partial_x \xi + v_y \partial_y \xi = - (U/2L) \phi$,
again using the approximations in Eqs.~(\ref{approx1}) and
(\ref{approx2}). Thus
$(L/U) d/dt = - (\phi/2)d/d \xi$ and Eqs.~(\ref{polx}) and
(\ref{poly}) become
\begin{eqnarray} \label{polxN} \phi \frac{d l_x}{d\xi} &=&
\frac{(1+a)(l_x-l_{0x})}{{\rm Wi}(1+aR)} +
\xi \phi_{\xi \xi} l_x - 2\sqrt{\rm Re} \phi_{\xi \xi} l_y \\
\phi \frac{d l_y}{d\xi} &=&  \frac{(1+a)(l_y-l_{0y})}{ {\rm Wi}(1+aR)} +
\frac{\xi^2 \phi_{\xi\xi}}{2 \sqrt{\rm Re}}  l_x -
\xi \phi_{\xi
\xi} l_y \label{polyN}
\end{eqnarray}

The quantity $R_{xy} $ is the average of $l_x l_y$ over all the
polymers contained in a small volume centered near the point
$(x,y)$, over the angle $\alpha$, and over the thermal noise. Such
an average would depend on the precise distribution of the
polymers, which is not obvious to obtain. Thus instead of
performing such an average, we calculate $l_x$ and $l_y$ for one
polymer for a fixed angle $\alpha = \pi/4$ such that
$l_{0x}=l_{0y}=l_0=1/\sqrt{2}$, and estimate $R_{xy}$ as some
function of the calculated $l_x$ and $l_y$. In an earlier short
communication~\cite{Letter}, we have shown that neglecting the
thermal noise and in the limit of small Wi and $a=0$, the
averaging over the angle $\alpha$ gives $R_{xy} =
[1+h(\xi)]\phi_{\xi \xi} {\rm Wi} \sqrt{\rm Re}$ with $h(\xi) \to
0$ as $\xi \to \infty$. We expect this behavior of $R_{xy}$ going
like $ {\rm Wi} \sqrt{\rm Re} \phi_{\xi \xi}$ far away from the
plate is generically true. Moreover, far away from the plate, $R
\to 1 $ and $l_y \to l_0$. Thus we estimate $R_{xy}$ as
\begin{equation}
R_{xy} =
{\rm Wi} \sqrt{\rm Re}\left(\frac{1+aR}{1+a} \right)
\phi_{\xi\xi} + (l_x-l_{0})(l_y-l_{0}) \label{rxy}
\end{equation}
which has the desired asymptotic behavior far away from the plate.
Defining
\begin{equation}
g(\xi) \equiv
\frac{[l_x(\xi)-l_{0}][l_y(\xi)-l_{0}](1+a)}{{\rm Wi}\sqrt{\rm
Re}[1+aR(\xi)]
\phi_{\xi\xi}(\xi)}
\label{g}
\end{equation}
with $R(\xi) = [l_x^2(\xi)+l_y^2(\xi)]^{1/2}$, then
\begin{equation}
R_{xy} = {\rm Wi} \sqrt{\rm Re}
\frac{(1+aR)}{(1+a)} \phi_{\xi\xi} (1+g)
\label{modelRxy}
\end{equation}
Substituting Eq.~(\ref{modelRxy})
into Eq.~(\ref{basic}), we have
\begin{equation}
2 \phi_{\xi\xi\xi} +
\phi \phi_{\xi\xi} + 2 \gamma \frac{d}{d \xi}
(g \phi_{\xi\xi}) = 0 \label{modify} \end{equation}

Comparing Eq.~(\ref{modify}) with Eq.~(\ref{BPphi}), it is obvious that
the effect of the polymers is contained in the term with $g(\xi)$.
Furthermore, this effect of the polymers
is equivalent to producing a space-dependent
viscosity $\nu_{\rm eff}(\xi)$. Specfically, if we define $\nu_{\rm
eff}$ by
\begin{equation}
\frac{\nu_p }{\tau(\xi)} R_{xy}(\xi) \equiv \nu_{\rm eff}(\xi) \partial_y v_x
\label{nueff}
\end{equation}
and take
\begin{equation}
\nu_{\rm eff}(\xi) = \nu_p [1+g(\xi)]
\end{equation}
then we obtain exactly Eq.~(\ref{modify}) from Eq.~(\ref{PBR}).
The total effective viscosity of the polymer solution is thus
\begin{equation}
\nu_{\rm tot}(\xi) = \nu + \nu_{\rm eff}(\xi) =  \nu_0 + \nu_p g(\xi)
\label{toteff}
\end{equation}
with the nontrivial effect of the polymers contained in the term
$\nu_p g(\xi)$, which is a function of the polymer concentration.

Equations~(\ref{polxN}), (\ref{polyN}), (\ref{g}) and (\ref{modify}) should be
solved consistently. With the obtained $\phi(\xi)$,
Eq.~(\ref{temperature}) is solved to obtain $\theta(\xi)$ and from which
Nu is obtained by using Eq.~(\ref{Nutheta}).
We would like to compare this Nu in the presence of polymers
to the reference value Nu$_0$
for a Newtonian
fluid with the \it{same} kinematic viscosity $\nu_0$ at the plate.
To get the reference value Nu$_0$, we start with the Prandtl-Blasius
velocity profile for a Newtonian fluid of kinematic viscosity $\nu_0$,
denoted as $\phi^{(B)}(\xi)$. This is just the solution
to Eq.~(\ref{BPphi}) with $\xi$ given by Eq.~(\ref{xinew})
or Eq.~(\ref{modify}) with $\gamma=0$.
With this $\phi^{(B)}$, we solve
Eq.~(\ref{temperature}) and obtain the resulting temperature profile
$\theta^{(B)}(\xi)$ at
the same Pr $=\nu_0/\kappa$ as the polymer solution. Then
\begin{equation}
{\rm Nu}_{0} = \sqrt{\frac{UL}{\nu_0}} [-\theta^{(B)}_\xi(0)]
\end{equation}
We are interested in the ratio
\begin{equation}
\frac{\rm Nu}{\rm Nu}_{0} =
\frac{\theta_\xi(0)} {\theta^{(B)}_\xi(0)}
\label{Nuration}
\end{equation}

\section{Calculations}
\label{cal}

We solve Eqs.~(\ref{polxN}), (\ref{polyN}), (\ref{g}) and (\ref{modify})
consistently by iteration at fixed values of Re and $a$.
We start with $\phi=\phi^{(B)}$
in Eqs.~(\ref{polxN}) and (\ref{polyN}) and solve for $l_x(\xi)$ and
$l_y(\xi)$. Using the boundary condition that
$l_x$ and $l_y$ $\to l_0$ as $\xi \to \infty$, we put
$l_x=l_y=l_0$ at some large value of $\xi$, denoted as $\xi_{\infty}$
(we use $\xi_{\infty}=30$) and
integrate backwardly until $\xi$ equals to some finite $\xi_0$ close to
zero. We cannot integrate forwardly in
$\xi$ because $\phi$ vanishes at
$\xi=0$~[see Eq.~(\ref{velBC})]. With the calculated $l_x$ and
$l_y$, we obtain $g(\xi)$ for $\xi$ between $\xi_0$ and $\xi_{\infty}$
using Eq.~(\ref{g}). From
Eq.~(\ref{polyN}), we see that $l_y(0)=l_0$ and thus $g(0)=0$.
Between $\xi=0$ and $\xi_0$ we
extrapolate $g(\xi)$ using a polynomial fit.
We input this $g$ into Eq.~(\ref{modify}) to solve for an updated $\phi$.
Then we use this updated $\phi$ in Eqs.~(\ref{polxN}) and (\ref{polyN})
to obtain an updated $g$. We repeat the procedure until
convergence in both $g$ and $\phi$ is achieved.

\begin{figure}
    \vspace{1cm}
       \centerline{\includegraphics[width=0.52\textwidth]{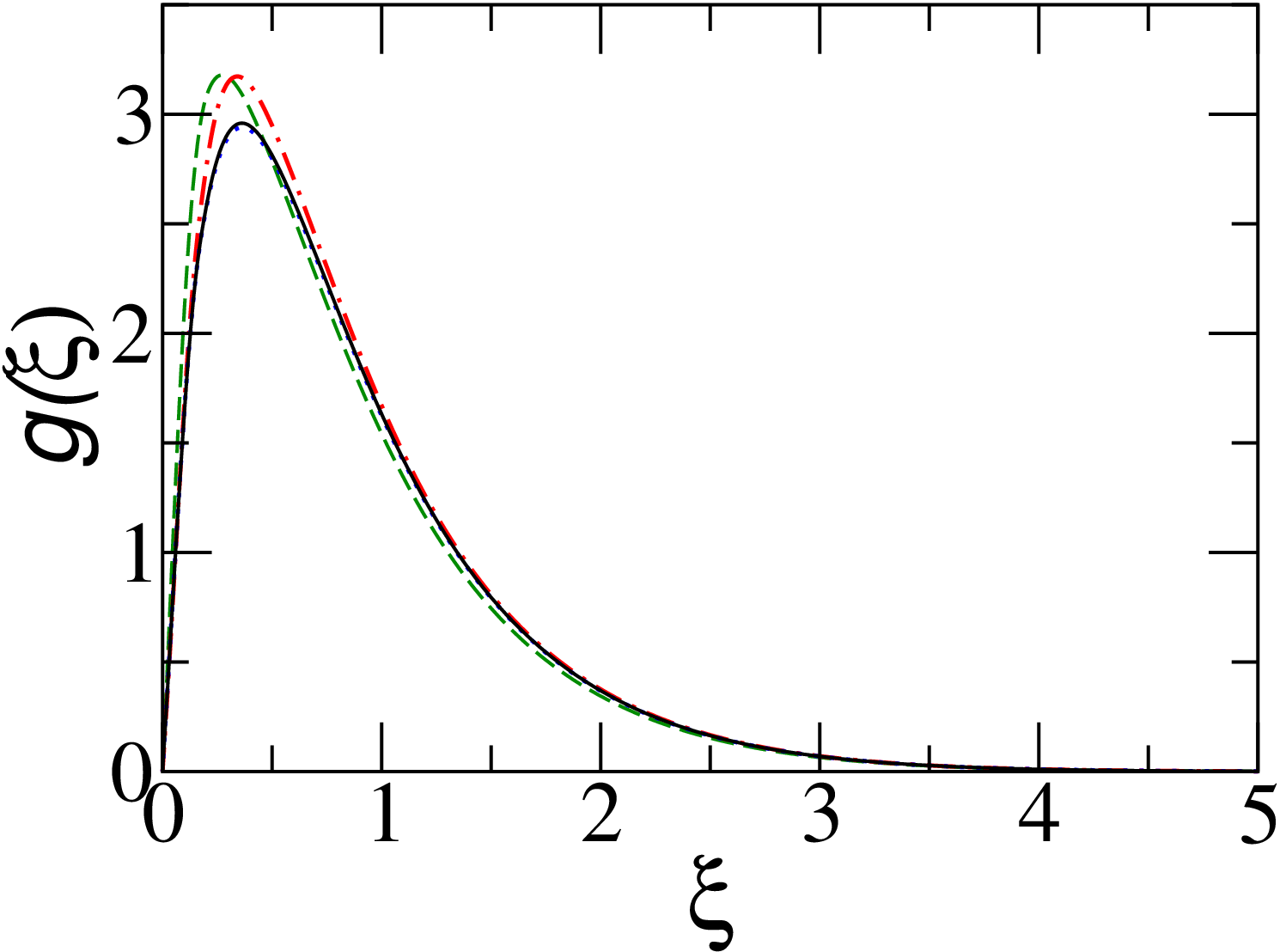}}
        \caption{Fast convergence of the numerical iterative procedure.
Result of $g(\xi)$ obtained after the first (dashed), second (dot-dashed),
fifth (dotted) and 10th iteration (solid).}
        \label{fig1}
\end{figure}

In Fig.~\ref{fig1}, we show $g(\xi)$ obtained using this iterative
procedure at Re$ = 4900$, $a=0.01$, Wi = 2.8 and $\gamma~=~0.2$.
Convergence is fast and achieved after only a few iterations.
Using the converged $\phi$, we obtain Nu for Pr=4.4 as discussed in
Section~\ref{problem}, and study the ratio Nu/Nu$_0$ for different
values of the parameters.

There is a constraint on the
possible values of the parameters imposed by the realizability of
$l_x(0)$. From Eq.~(\ref{polxN}),
we see that $l_x(0)$ satisfies the following equation:
\begin{equation}
\frac{1+1/a}{2{\rm Wi} \sqrt{\rm Re} \phi_{\xi\xi}(0) l_0}
[l_x(0)-l_0] = \sqrt{l_x(0)^2+l_0^2} + \frac{1}{a}
\label{lx0}
\end{equation}
Thus the condition for Eq.~(\ref{lx0}) to have a finite real solution
for $l_x(0)$ is:
\begin{equation}
\left(1+\frac{1}{a}\right) \frac{1}{\rm Wi}
> 2\sqrt{\rm Re} \phi_{\xi\xi}(0) l_0
\label{constraint}
\end{equation}
As a result,
for a given value of Re,
the range of allowed values of Wi is smaller for
larger values of $a$. Moreover, the allowed range of Wi is larger for
a smaller value of Re. We have used two values of Re: 100 and
4900.
For Re=4900, we take $a=0.01$ and find that for $\gamma=0.2$,
the maximum allowed value
of Wi is about 2.8. For Re=100, we take $a=0.07$ and study different
values of Wi up to
3.0 for the same value of $\gamma$. We then fix Wi=2.5, and study the
effect of polymer concentration by varying
the value of $\gamma$ for Re=4900 and $a=0.01$.
We study also a few other values of the parameters to investigate
the dependence of the results on the parameters.

\section{Results and Discussions}
\label{result}

\begin{figure}
    \vspace{1cm}
        \begin{center}
        \includegraphics[width=0.5\textwidth]{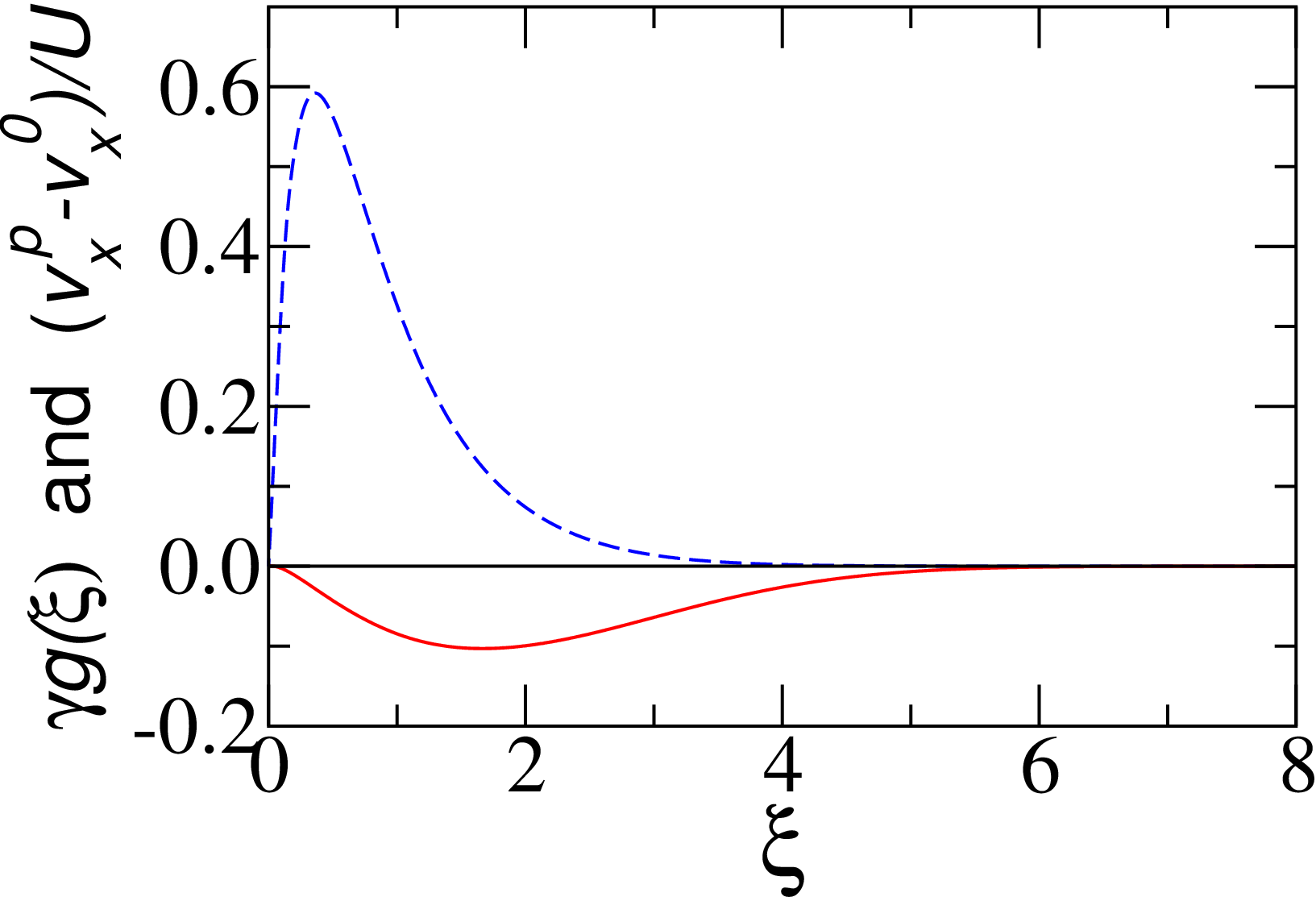}
        \end{center}
        \caption{Effective viscosity due to the
polymers as represented by $\gamma g(\xi)$~(long dashed curve) and the
difference in the horizontal velocity of the polymer solution from that of the pure
solvent, $(v_x^p-v_x^0)/U$~(solid curve),
as a function of $\xi$ for Re=4900, $a=0.01$, Wi=0.28, and $\gamma=0.2$.}
        \label{fig2}
\end{figure}

We find that
$g(\xi)$ increases from zero at the plate ($\xi=0$)
up to a certain maximum then
decreases rather rapidly back to zero far away from the plate ($\xi \to
\infty$). Whenever $g(\xi)$ is larger than zero, the viscosity of the
polymer solution is enhanced compared to that of the solvent.
We expect that such an increase in the viscosity
gives rise to an enhancement of the friction drag,
which would in turn
result in a reduction of the horizontal velocity. Indeed, we find
that the horizonal velocity, $v_x^p$ for the flow with polymers
is reduced compared to $v_x^0$, the horizontal velocity for the flow without
polymers~(see Fig.~\ref{fig2}).

\begin{figure}
    \vspace{1cm}
        \begin{center}
        \includegraphics[width=0.5\textwidth]{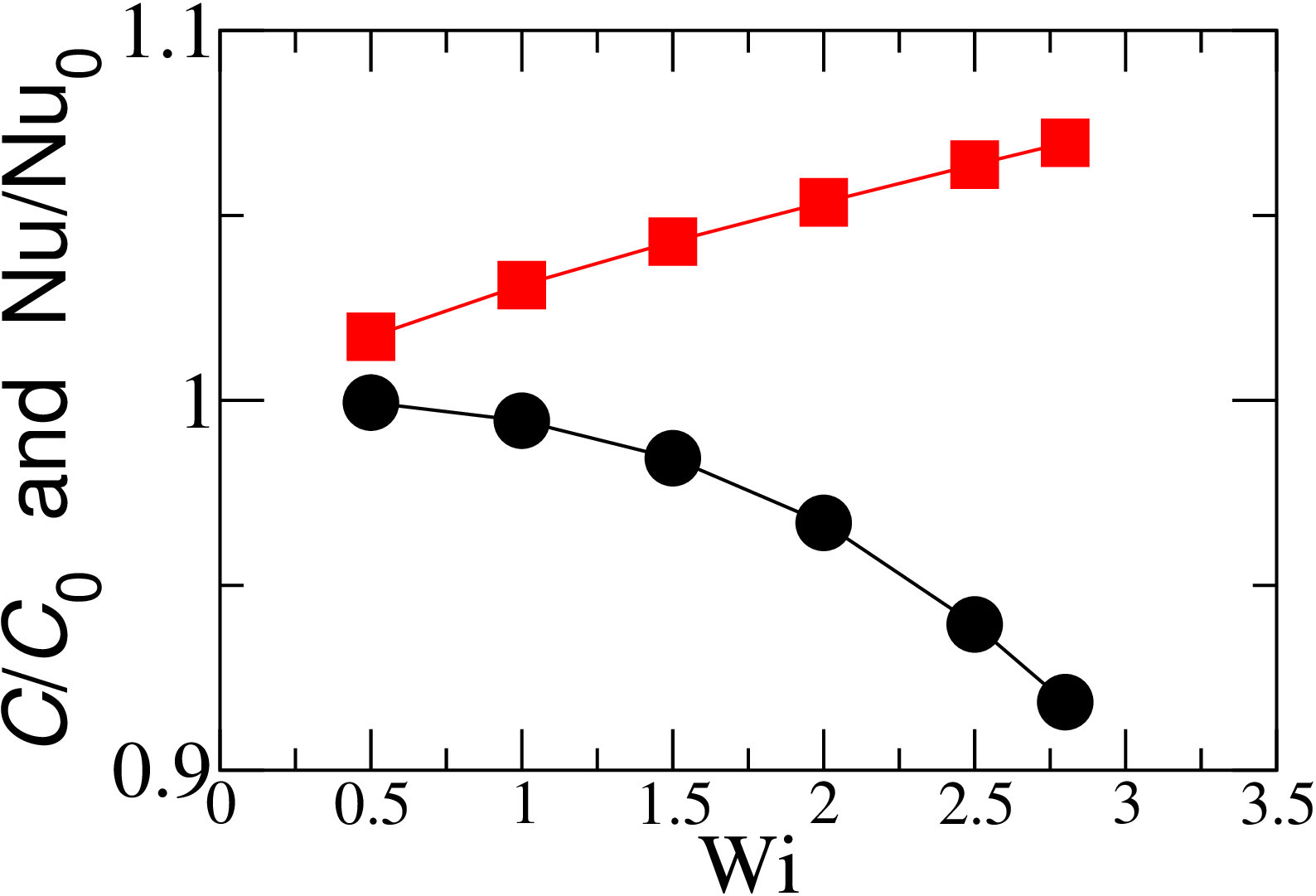}
        \end{center}
        \caption{$C/C_0$~(squares) and Nu/Nu$_0$~(circles) as a function
of Wi for Re=4900, $a$=0.01 and $\gamma=0.2$.}
        \label{fig3}
\end{figure}

We study the change in drag by measuring directly
the drag coefficient, defined by:
\begin{equation}
C = \frac{\nu_0 \partial_y v_x \big|_{y=0}}{1/2 U^2} =
\frac{1}{2 \sqrt{\rm Re}}
\phi_{\xi \xi}(0)
\label{drag}
\end{equation}
We are interested in the ratio
\begin{equation}
\frac{C}{C_0} =
\frac{\phi_{\xi\xi}(0)}{\phi^{(B)}_{\xi\xi}(0)}
\label{ratio2}
\end{equation}
An enhancement in the friction drag implies a
reduction of heat transport. To see this,
we note that upon double integration of
Eq.~(\ref{temperature}) by $\xi$, we obtain:
\begin{equation}
\label{dtheta}
-\theta_\xi(0) = \frac{1}{ \int_{0}^{\infty}
d t \exp[-\frac{\rm Pr}{2} \int_{0}^{t}\phi(s) ds]}
\end{equation}
which tells us that Nu is a functional of $\phi$.
Define $\Phi(\xi) \equiv \int_0^\xi \phi(s)ds$, we can calculate
$\delta {\rm Nu}$, the variation in Nu due to a variation in $\Phi$:
\begin{equation}
\label{carina}
\delta {\rm Nu} = \frac{{\rm Pr}}{2 \sqrt{\rm Re}}
{\rm Nu}^2 \int_0^\infty \exp(-{\rm Pr}
\Phi(s) /2) \delta \Phi (s) ds
\end{equation}
In the above expression,
 $\delta \Phi$ represents the variation of $\Phi$ due to the effect of the polymers.
Since $v_x = U \phi_{\xi}$, the mass throughput in the $x$ direction across a distance
$\xi$ is
given by $U \phi(\xi)$, and thus a
reduction in mass throughput implies an $\delta \Phi < 0$.
Therefore, Eq.~(\ref{carina})
shows that drag enhancement, i.e., reduction in the mass throughput implies a reduction of Nu.
It is indeed found that $C/C_0 > 1$ while Nu/Nu$_0 < 1$~(see
Fig.~\ref{fig3}).

\begin{figure}
\vspace{1cm}
        \begin{center}
        \includegraphics[width=0.5\textwidth]{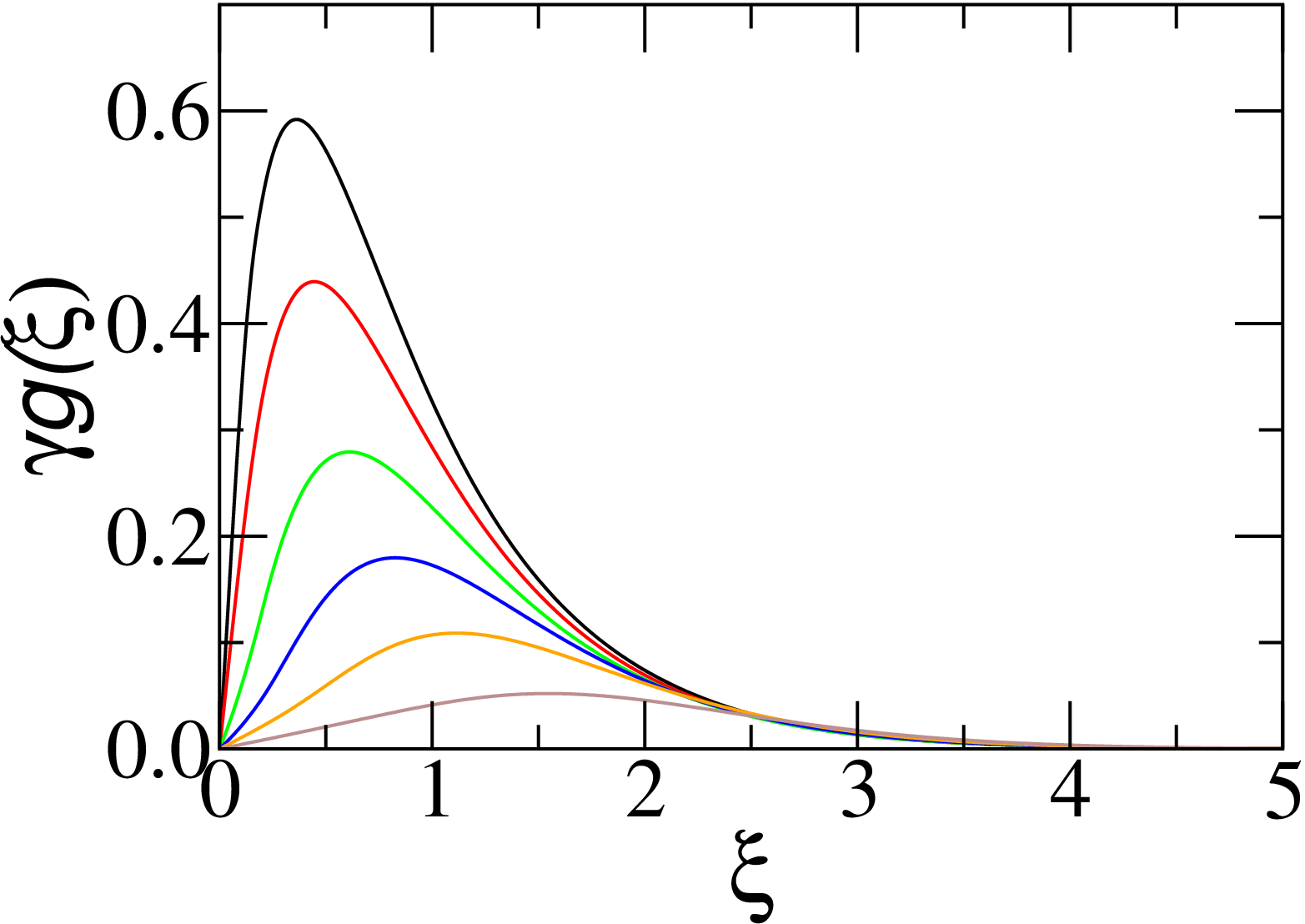}
        \end{center}
        \caption{Dependence of the effective viscosity as measured by
$\gamma g(\xi)$ on Wi at fixed
polymer concentration for Re=4900, $a$=0.01 and $\gamma=0.2$. From
bottom to top, Wi increases from 0.5, 1.0, 1.5, 2.0. 2.5 to 2.8.}
        \label{fig4}
\end{figure}

As shown in Fig.~\ref{fig3}, the amount of drag enhancement and heat
reduction increases with Wi. This is understood as the result of an
increase in the effective viscosity with Wi. In
Fig.~\ref{fig4}, we show the dependence of $g(\xi)$ on Wi.
It can be seen that
the effective viscosity increases with Wi while
the region
in which the polymers are active (i.e., $g(\xi)>0$) is approximately
independent of Wi.

\begin{figure}
\vspace{0.8cm}
        \begin{center}
        \includegraphics[width=0.5\textwidth]{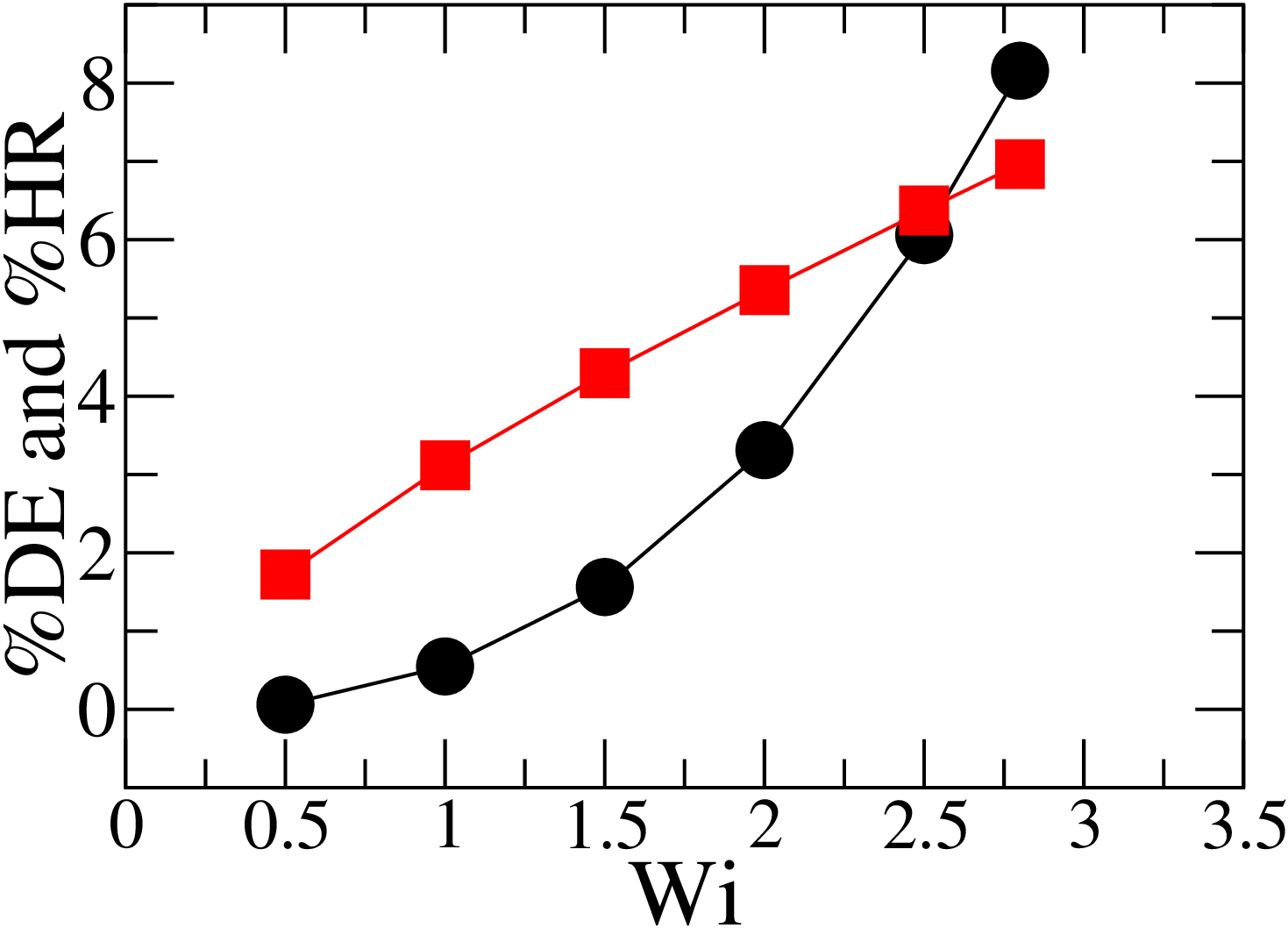}
        \end{center}
        \caption{Percentage of drag enhancement (\% DE)~(squares) and
heat reduction (\%HR)~(circles) as a function of Wi at fixed
$\gamma=0.2$ for Re=4900 and $a$=0.01.}
        \label{fig5}
\end{figure}

\begin{figure}
\vspace{0.8cm}
        \begin{center}
        \includegraphics[width=0.5\textwidth]{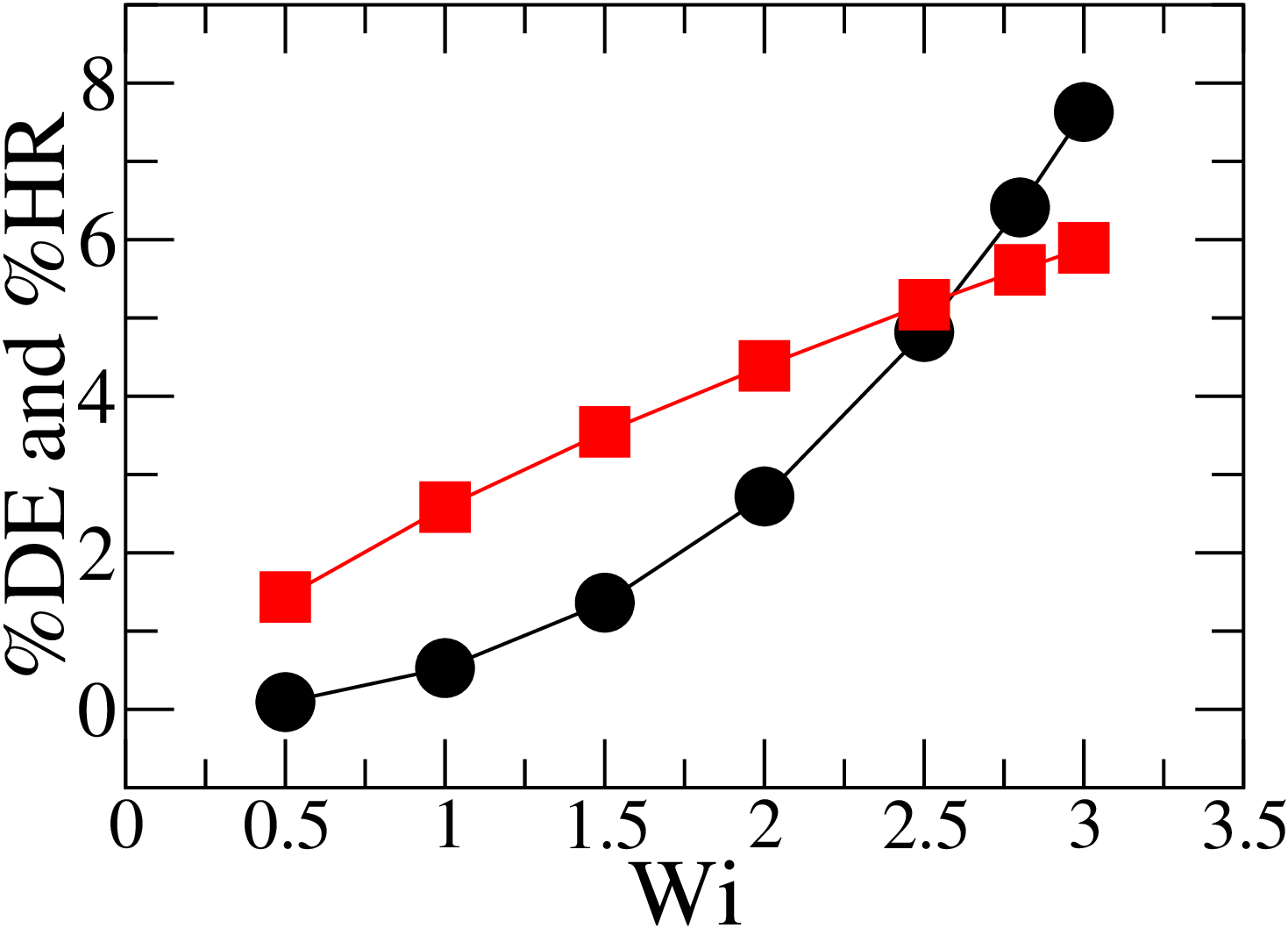}
        \end{center}
        \caption{\%DE (squares) and \%HR~(circles)
as a function of Wi at fixed $\gamma=0.2$ for Re=100 and $a=0.07$.}
        \label{fig5add}
\end{figure}

In Figs.~\ref{fig5} and \ref{fig5add}, we show the dependence of the
amount of drag enhancement (DE), $C/C_0-1$,
and heat reduction (HR), $1-{\rm Nu}/{\rm Nu}_0$, as
a function of Wi at a fixed polymer concentration for the two
different sets of values of Re and $a$ studied.
It can be seen that both effects are relatively modest.
The effect is generally larger when Re is larger.
Both \%DE and \%HR increases as Wi
increases and for the range of Wi studied, they increase
quadratically with Wi. Interestingly, for \%DE
the rate of increase decreases with Wi and this leads to the possibility
of the saturation of the effect at large Wi. On the other hand, for
\%HR, the rate of increase increases with Wi thus the effect on heat
reduction is larger than that  but increases for \%HR.

\begin{figure}
    \vspace{1cm}
        \begin{center}
        \includegraphics[width=0.5\textwidth]{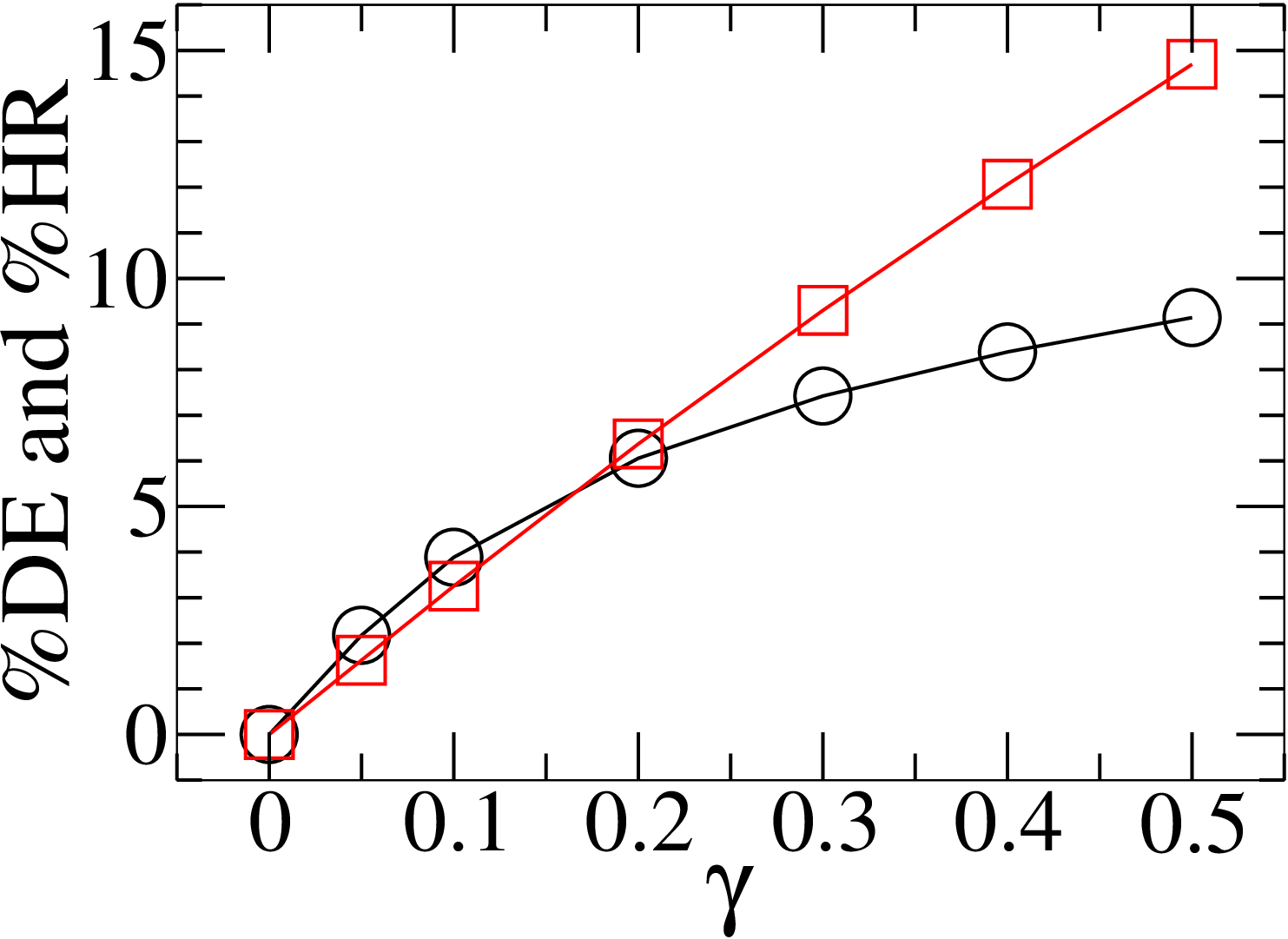}
        \end{center}
        \caption{Dependence of the effect on polymer concentration:
\%DE~(squares) and \%HR~(circles) as a function of $\gamma$ at fixed Wi=2.5.}
        \label{fig6}
\end{figure}

\begin{figure}
    \vspace{1cm}
        \begin{center}
        \includegraphics[width=0.5\textwidth]{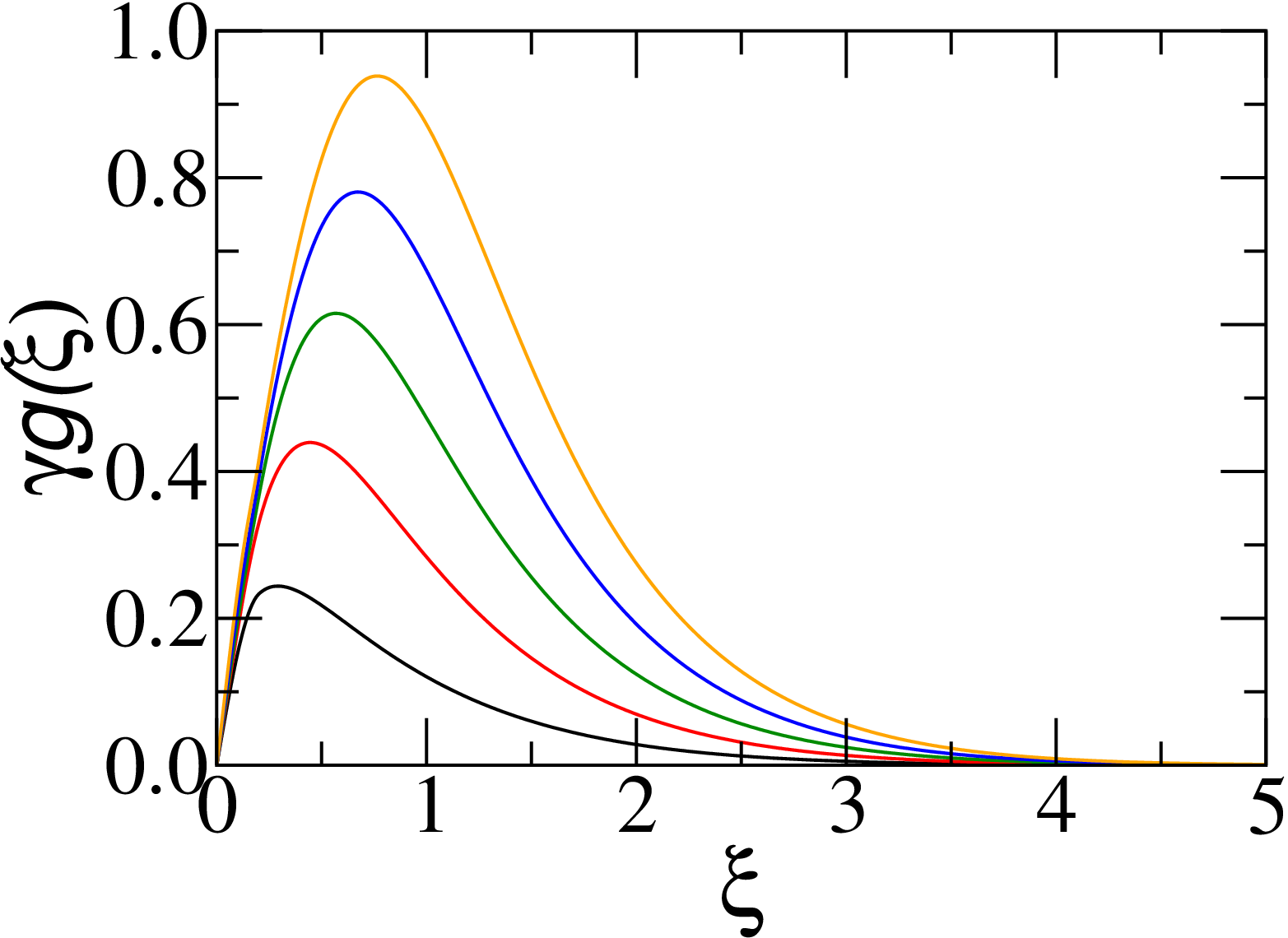}
        \end{center}
        \caption{Dependence of $\gamma g(\xi)$ on $\gamma$ at fixed Wi=2.5. From
bottom to top, $\gamma$ increases from 0.1 to $0.5$ in steps of 0.1.}
        \label{fig7}
\end{figure}

In Fig.~\ref{fig6}, we fix Re=4900, a= 0.01, Wi=2.5 and show the
extent of drag enhancement and heat reduction as a function of polymer
concentration specified by $\gamma$. We see that the effect increases
with $\gamma$ as expected. Moreover, we see a possible saturation of the effect
in the limit of the large polymer concentration.
In Fig.~\ref{fig7}, we show the dependence of $\gamma g(\xi)$ on $\gamma$.
Again we see that $\gamma g$ increases with $\gamma$.
It is this increase of the effective viscosity
of the polymers with
polymer concentration that
leads to the increase in the extent of drag enhancement and heat
reduction when polymer concentration increases.

We summarize the results for \%DE and \%HR obtained for different values of the
parameters in Table~1.
We see that the effect generally increases with
$a$, Re, Wi and $\gamma$ while the other parameters are kept fixed.
We note that drag enhancement and heat reduction
are found even for $a=0$ although the effect is relatively modest with
only a few percentage of HR for $\gamma=0.5$ and Wi=2.5. The effect
increases with $a$ and interestingly the increase in \%HR is larger
than that the increase in \%DE when $a$ is increased.

\newpage
\begin{table}
\begin{center}
\begin{tabular}{|c|c|c|c|c|c|} \hline
Re & $a$ & Wi & $\gamma$ & \% HR & \% DE \\ \hline
100 & 0.005 & 3 & 0.2 & 2.1 & 4.5  \\
100 & 0.05 & 3 & 0.2 & 5.0 & 5.4  \\
100 & 0.2 & 0.5 & 0.2 & 0.2 & 1.6 \\
100 & 0.4 & 0.5 & 0.2 & 0.5 & 2.0 \\
100 & 0.6 & 0.5 & 0.2 & 0.9  & 2.4 \\
100 & 0.7 & 0.5 & 0.2 & 1.3 & 2.5 \\ \hline
100 & 0.07 & 1.0 & 0.2 & 0.5 & 2.6 \\
100 & 0.07 & 1.5 & 0.2 & 1.4 & 3.5 \\
100 & 0.07 & 2.0 & 0.2 & 2.7 & 4.4 \\
100 & 0.07 & 2.5 & 0.2 & 4.8 & 5.2 \\
100 & 0.07 & 2.8 & 0.2 & 6.4 & 5.6 \\
100 & 0.07 & 3 & 0.2 & 7.6 & 5.9  \\ \hline
4900 & 0.0 & 2.5 & 0.5 & 3.0 & 11.2 \\
4900 & 0.0 & 2.5 & 0.2 & 1.5 & 4.7 \\
4900 & 0.0 & 2.8 & 0.2 & 1.8 & 5.0 \\
4900 & 0.1& 0.3 & 0.2 &  0.7 & 2.2 \\
4900 & 0.05 & 0.6 & 0.2 & 1.7  & 3.4 \\
4900 & 0.005 & 3 & 0.2 & 4.2 & 6.1 \\
4900 & 0.005 & 4 & 0.2 & 7.8 & 7.3 \\
4900 & 0.005 & 5 & 0.2 & 12.1 & 8.4 \\ \hline
4900 & 0.01 & 1.0 & 0.2 & 0.6 & 3.1 \\
4900 & 0.01 & 1.5 & 0.2 & 1.6 & 4.3 \\
4900 & 0.01 & 2.0 & 0.2 & 3.3 & 5.4 \\
4900 & 0.01 & 2.5 & 0.2 & 6.1 & 6.3 \\
4900 & 0.01 & 2.8 & 0.2 & 8.2 & 7.0 \\ \hline
4900 & 0.01 & 2.5 & 0.05 & 2.2 & 1.6 \\
4900 & 0.01 & 2.5 & 0.1 & 3.9 & 3.2 \\
4900 & 0.01 & 2.5 & 0.3 & 7.4 & 9.3 \\
4900 & 0.01 & 2.5 & 0.4 & 8.4 & 12.1 \\
4900 & 0.01 & 2.5 & 0.5 & 9.1 & 14.7 \\ \hline
\end{tabular}
\caption{Amount of drag enhancement and heat reduction for different
values of parameters.}
\label{table1}
\end{center}
\end{table}

\vspace{1cm}

\section{Summary and Conclusions}
\label{conclusion}

In this paper, we have studied the problem of heat transport in
steady-state Prandtl-Blasius flow with polymers near a slightly heated
plate. We have shown how a set of equations can be written for the
problem and how this set of equations can be solved numerically by an
iterative procedure. Our results demonstrate that the physical effect of
the polymers is equivalent to producing a space-dependent
effective viscosity, which increases near the plate.
Because of this increase in viscosity, drag is enhanced. We have shown
that such a drag enhancement then leads to a reduction in heat
transport in the Prandtl-Blasius flow with polymers.
As discussed in Sec.~\ref{introduction}, in turbulent RB convection of Ra about
$10^{10}$, heat transport is dominated by contributions from the
boundary layers and the mean velocity and temperature
boundary layer profiles were found to be well-described
by the Prandtl-Blasius profiles. Hence, our theory
may explain the recent experimental observation of a reduction in heat
transport in turbulent RB convection with polymer additives.
In particular, an amount of about 10 \% HR can be obtained
with suitable parameters in our theory, and this amount of heat
reduction is comparable to that observed in the experiment.

\bigskip

The work of ESCC and VWSC was supported in part by the Hong Kong Research Grants Council
(CUHK 400708).

\end{document}